\documentclass[12pt]{iopart}
\pdfoutput=1
\usepackage[utf8]{inputenc} 			
\usepackage[T1]{fontenc}				
\usepackage{lmodern}					
\usepackage[pdftex]{graphicx} 		    
\usepackage[svgnames]{xcolor}
\usepackage{cite}
\usepackage{iopams}                     
\usepackage{setstack}                   
\usepackage[english]{babel}
\usepackage{lmodern}

\usepackage{ulem}

\usepackage{listings}
\usepackage{tikz}
\graphicspath{{./}{./Images/}}
\newcommand{\eqref}[1]{\eref{#1}}

\newcommand{\cP}{\mathcal{P}}
\newcommand{\cL}{\mathcal{L}}

\definecolor{linkcolor}{rgb}{0,0,0.6}

\usepackage[pdftex,colorlinks=true,pdfstartview=FitV,linkcolor= linkcolor,citecolor= linkcolor,urlcolor= linkcolor,hyperindex=true,hyperfigures=false]{hyperref}
\begin{document}

\title[AOUP in the presence of Brownian noise: a perturbative approach]{AOUP in the presence of Brownian noise: a perturbative approach}
\author{David Martin, Thibaut Arnoulx de Pirey}
\address{Laboratoire Matière et Systèmes Complexes,
              UMR 7057 CNRS/P7,
              Université de Paris,
              10 rue Alice Domon et Léonie Duquet,
              75205 Paris Cedex 13,
              France}
\ead{david.martin91120@gmail.com}
\date{\today\ -- \jobname}

\begin{abstract}
By working in the small persistence time limit, we determine the steady-state distribution of an Active Ornstein Uhlenbeck Particle (AOUP) experiencing, in addition to self-propulsion, a Gaussian white noise modelling a bath at temperature $T$.
This allows us to derive analytical formulas for three quantities: the spatial density of a confined particle, the current induced by an asymmetric periodic potential and the entropy production rate.
These formulas disentangle the respective roles of the passive and active noises on the steady state of AOUPs, showing that signatures of non-equilibrium can display surprising behaviors as the temperature is varied.
Indeed, depending on the potential in which the particle evolves, both the current and the entropy production rate can be non-monotonic functions of $T$. The latter can even diverge at high temperature for steep enough confining potentials.
Thus, depending on context, switching on translational diffusion may drive the particle closer to or further away from equilibrium.
We then probe the range of validity of our quantitative derivations by numerical simulations.
Finally, we explain how the method presented here to tackle perturbatively an Ornstein Uhlenbeck (OU) noise could be further generalized beyond the Brownian case.\\

\end{abstract}
\submitto{\JSTAT}
\section{Introduction}
One of the challenges in the field of Active Matter is to understand and predict the emerging properties of assemblies composed of individual agents able to produce mechanical work by locally dissipating energy \cite{marchetti2013hydrodynamics}.
By their very nature, such systems break detailed balance and thus lie within the realm of out-of-equilibrium physics.
At a theoretical level, the modelling of the individual units of active systems reveals their nonequilibrium nature by involving non-Gaussian and colored noises differing from the familiar Wiener process of thermal-equilibrium physics.
On the one hand, this approach has been fruitful and both analytical and numerical works have shown that intriguing phenomena, prohibited in equilibrium physics, arise for such noises : accumulation near repulsive walls \cite{elgeti2009self,tailleur2009sedimentation,caprini2018active}, emergence of currents in asymmetric periodic potentials \cite{CurrentGreek,lindner1999inertia,sandford2017pressure,magnasco1993forced,bartussek1997ratchets,koumakis2014directed}, collective motion \cite{vicsek1995novel,gregoire2004onset}, motility-induced phase separation \cite{tailleur2008statistical,cates2015motility}...
On the other hand, from a theoretical standpoint, resorting to these unusual noises has the technical disadvantage of making algebraic manipulations more cumbersome. Quantifying the departure from equilibrium physics and determining the corresponding steady-state distribution is a whole research field \textit{per se} \cite{bianucci2020optimal,szamel2014self,solon2015active,basu2019long,basu2020exact,hartmann2020convex,shee2020active,caprini2019activity}.
In this article, we focus on an Active Ornstein Uhlenbeck Particle (AOUP) evolving in one space dimension, subjected to an external potential $\phi(x)$ and further experiencing an additional thermal noise. Its position $x(t)$ and self-propulsion $v(t)$ evolve according to the following system of Langevin equations \cite{goswami2020work,dabelow2020irreversible}:
\begin{eqnarray}
  \label{eq:langevin_1}
  \dot{x}&=&-\partial_x \phi + \sqrt{2T}\ \eta_1 + v \\
  \label{eq:langevin_2}
  \dot{v}&=&-\frac{v}{\tau} + \frac{\sqrt{2D}}{\tau}\ \eta_2\ .
\end{eqnarray}
In the above dynamics \eref{eq:langevin_1}-\eref{eq:langevin_2}, $\eta_1$ and $\eta_2$ are two uncorrelated Gaussian white noises of unit variance, $T$ is the amplitude of the thermal noise while $D$ and $\tau$ control the amplitude and the persistence of the self-propulsion.
When $T=0$, \eref{eq:langevin_1}-\eref{eq:langevin_2} correspond to the workhorse AOUP model which has been used to model transport properties of active colloids \cite{koumakis2014directed} as well as collective cell dynamics \cite{hakim2017collective,deforet2014emergence}.
On the theoretical side, there has been fundamental interest in its steady-state distribution, which has been characterized both in the limit of small $\tau$ \cite{FodorAOUP,bonilla2019active,martin2020aoup,woillez2020nonlocal,jung1987dynamical} and in the limit of high $\tau$ \cite{fily2019self,woillez2020nonlocal,jung1987dynamical}.
However, these theoretical approaches ignore the physically relevant presence of an underlying thermal noise and the steady state distribution of \eref{eq:langevin_1}-\eref{eq:langevin_2} remains elusive for a generic combination of $T$ and $D$.
Indeed, such a combination of both active and thermal noise sources arises in multiple experiments: passive tracers embedded in living cells \cite{ wilhelm2008out, fodor2015activity, ahmed2018active} or immersed in a bath of active colloids \cite{maggi2014generalized}, flucutations of cellular membrane in red blood cells \cite{turlier2016equilibrium, ben2011effective}...
\par
In this article, we aim at filling this gap by computing perturbatively the stationary probability density of an AOUP experiencing an additional thermal noise in the small-persistence-time limit.
Note that for $\tau=0$, the self-propulsion $v$ falls back onto a Wiener process of amplitude $D$. In this particular case, the dynamics \eref{eq:langevin_1}-\eref{eq:langevin_2} is an equilibrium one with temperature $T+D$.
Thus, intuitively, one could hope to find analytical formulas that smoothly departs from thermal equilibrium when $\tau$ is small.
We develop here such a perturbative expansion and our main result is an analytical prediction of the steady-state distribution $\cP_s(x,v)$ as a series in $\tau^{1/2}$.
Building on it, we make quantitative predictions about three emerging quantities: the marginal in space of the probability density, the current in an asymmetric periodic ratchet, and the entropy production rate.
Depending on the boundary conditions and on the potential $\phi(x)$, we find that the interplay between passive and active noises leads to a rich phenomenology for the current and the entropy production rate when the temperature is varied: decline or non-monotonicity, divergence or decay at high $T$.
\section{Systematic construction of the probability density function}
\label{sec:pdf}
To perform the derivation of the steady-state distribution $\cP_s(x,v)$ as a series in powers of $\tau^{1/2}$, we proceed in several steps as follows. First, we conveniently rescale the Fokker-Planck operator. Then, we look for its stationary solution by expanding $\cP_s$ on the basis of Hermite polynomials and we show how the Fokker-Planck equation imposes a recurrence relation between the coefficients of this expansion. Finally, we solve the recurrence by expanding these coefficients as power series in $\tau^{1/2}$.
We now detail the derivation starting from the Fokker-Planck operator $\cL$ corresponding to \eref{eq:langevin_1}-\eref{eq:langevin_2}, which reads
\begin{equation}
\cL = \partial_{x}(\partial_{x}\Phi ) - v \partial_{x} + \partial_{v}\left(\frac{v}{\tau}\right)+\frac{D}{\tau^{2}}\partial_{vv} +T\partial_{xx}\ .
\end{equation}
Because the steady-state distribution of \eref{eq:langevin_2} is  proportional to $\exp(-\frac{\tau v^2}{2D})$, we now rescale $v$ as $\tilde{v}=\sqrt{\tau}v$ in order to expand $\cP_s$ in series of $\tau^{1/2}$ around the equilibrium measure.
Expressed in terms of the rescaled variable, $\mathcal{P}_s(x,\tilde{v})$ satisfies
\begin{equation}
  \label{eq:stationary_condition}
 \tilde{\mathcal{L}} \mathcal{P}_s(x,\tilde{v})=0
\end{equation}
with the operator $\tilde{\cL}$ defined as :
\begin{equation}
\label{eq:FP_operator}
\tilde{\mathcal{L}}= \partial_{x}(\partial_{x}\Phi ) - \frac{\tilde{v}}{\sqrt{\tau}} \partial_{x} + \partial_{\tilde{v}}\left(\frac{\tilde{v}}{\tau}\right)+\frac{D}{\tau}\partial_{\tilde{v}\tilde{v}}+T\partial_{xx}\ .
\end{equation}
In the remainder of this work, the tilde notation for $v$ and $\cL$ will be omitted for notational simplicity.
We first note that the Fokker-Planck operator \eref{eq:FP_operator} can be written as :
\begin{equation}
\label{eq:FP_operator_2}
\mathcal{L} = \frac{1}{\tau}\mathcal{L}_{1} + \frac{1}{\sqrt{\tau}} \mathcal{L}_{2} + \mathcal{L}_{3} \ ,
\end{equation}
where $\cL_1$, $\cL_2$ and $\cL_3$ are given by
\begin{equation}
\mathcal{L}_{1} = D\frac{\partial^{2}}{\partial^{2}v}+ \frac{\partial }{\partial v}v \qquad
\mathcal{L}_{2} = -v\frac{\partial }{\partial x} \qquad
\mathcal{L}_{3} = \frac{\partial }{\partial x}\frac{\partial \phi}{\partial x} + T\frac{\partial^2}{\partial^2 x}\ .
\end{equation}
$\mathcal{L}_{1}$ is the Fokker-Planck generator of the Ornstein-Uhlenbeck process, and its $n^{\rm th}$ eigenfunction $P_n$ is related to the $n^{\rm th}$ physicists' Hermite polynomial $H_n(v)=(-1)^n e^{v^2}\partial^n_v e^{-v^{2}}$ :
\begin{eqnarray}
P_{n}(v)&=&\frac{e^{-\frac{v^{2}}{2D}}H_{n}\left(\frac{v}{\sqrt{2D}}\right)}{\sqrt{2^{n}n!2\pi D}}\ .
\end{eqnarray}
The family $\{P_{n}\}$ are eigenfunctions of the operator $\cL_1$ satisfying
\begin{eqnarray}
\label{eq:eigenvalues}
\mathcal{L}_{1}P_{n} &= -n P_{n}
\end{eqnarray}
and they are further orthogonal to the family $\{H_n\}$ as
\begin{eqnarray}
\label{eq:orthogonality}
\delta_{k,n} &= \int_{-\infty}^{+\infty}\frac{ H_{k}\left(\case{v}{\sqrt{2D}}\right)}{\sqrt{2^k k!}}P_{n}(v)dv\ .
\end{eqnarray}
We use the $P_n$'s to search for the solution of the stationnary distribution $\cP_{s}$ under the form of:
\begin{eqnarray}
  \label{eq:ansatz_1}
\cP_{s}(x,v)&=&\sum_{n}P_{n}(v)A_{n}(x)\ .
\end{eqnarray}
Using the orthogonality property \eref{eq:orthogonality}, the $A_{n}$'s can be obtained as
\begin{eqnarray}
A_{n}(x)&=&\int \mathcal{P}_{s}(x,v)\frac{H_{n}\left(\case{v}{\sqrt{2D}}\right)}{\sqrt{2^{n} n!}}dv\ .
\end{eqnarray}
Inserting \eref{eq:ansatz_1} into \eref{eq:stationary_condition} and using \eref{eq:eigenvalues}, we find that $A_{n}$ is a solution of
\begin{eqnarray}
\fl
\label{eq:FP_non_projected}
\sum_{n}P_{n}(v)\partial_{x}\left(\partial_{x}\phi A_{n}\right) + \sum_{n}P_n(v)T\partial_{xx}A_{n}-\sum_{n}\frac{nP_{n}(v)}{\tau}A_{n}-\sum_{n}\frac{vP_{n}(v)}{\sqrt{\tau}}\partial_{x}A_{n}
=0\ .
\end{eqnarray}
Using the recurrence property of Hermite polynomials, $H_{n+1}(v)=2vH_n(v)-2nH_{n-1}(v)$, we decompose $vP_{n}$ into a sum of $P_{n+1}$ and $P_{n-1}$
\begin{eqnarray}
vP_{n}=\sqrt{(n+1)D}P_{n+1}+\sqrt{nD}P_{n-1}\ .
\end{eqnarray}
We are now in position to project equation (\ref{eq:FP_non_projected}) onto $H_{k}$ and use the orthogonality relation \eref{eq:orthogonality}.
This leads us to the following recursion relation for the $A_{n}$'s
\begin{eqnarray}
  \fl
  \label{eq::recurrence::eigenfunctions}
  0=-nA_{n}-\sqrt{\tau}\sqrt{(n+1)D}\partial_{x}A_{n+1}
  -\sqrt{\tau}\sqrt{nD}\ \partial_{x}A_{n-1} + \tau\partial_{x}\left(\partial_{x}\phi A_{n}\right)+\tau T\partial_{xx}A_n\ .
\end{eqnarray}
We now look for the $A_n$'s as series in powers of $\tau^{1/2}$.
Because \eref{eq:FP_operator} is formally invariant upon the reversal $\{\tilde{v},\sqrt{\tau}\}\to - \{\tilde{v},\sqrt{\tau}\}$, so is the stationary distribution $\cP_s$. Consequently, $A_{2k}$ contains only integer powers of $\tau$ while $A_{2k+1}$ contains only half-integer powers of $\tau$.
We shall further assume that the first nonzero contribution to $A_{k}$ is of order $\tau^{k/2}$. This hierarchical ansatz is necessary to disentangle and solve, starting from $A_0$ and order by order in powers of $\tau^{1/2}$, the recurrence equation \eqref{eq::recurrence::eigenfunctions}.
Its validity is a posteriori confirmed by inserting our final result for $\cP_s$ into \eqref{eq:FP_operator} and checking that $\mathcal{L}\cP_s$ vanishes order by order in $\tau$.
We thus propose the scaling ansatz
\begin{eqnarray}
\label{eq:scaling_A_0}
A_{0} &=& A_{0}^{0}(x)+\tau A_{0}^{2}(x)+\tau^2 A_{0}^{4}(x)+... \\
A_{1} &=& \tau^{1/2}A_{1}^{1}(x)+\tau^{3/2}A_{1}^{3}(x)+\tau^{5/2}A_{1}^{5}(x)+... \\
A_{2} &=& \tau A_{2}^{2}(x)+\tau^{2}A_{2}^{4}(x)+\tau^{3}A_{2}^{6}(x)+... \\
\nonumber &\vdots
\end{eqnarray}
Let us now show that the $A_{i}^{j}$ can be computed recursively. Looking at \eref{eq::recurrence::eigenfunctions} for $n=0$, we get
\begin{equation}
  \label{eq:A_1_int}
\partial_{x}A_{1}=\sqrt{\frac{\tau}{D}}\left[\partial_{x}\left(\partial_{x}\phi A_{0}\right) + T \partial_{xx}A_0\right]\ .
\end{equation}
Equating coefficients of order $\tau^{k/2}$ on both sides of \eref{eq:A_1_int} and integrating once over the position leads to:
\begin{equation}
  \label{eq:A_1_full}
A_{1}^{k}=\frac{1}{\sqrt{D}}\left[\partial_{x}\phi A_{0}^{k-1}+T\partial_x A_0^{k-1}\right] + b_k \ .
\end{equation}
with $b_k$ an integration constant. Further equating coefficients of order $\tau^{k/2}$ in \eref{eq::recurrence::eigenfunctions}, we obtain:
\begin{eqnarray}
  \fl
  \label{eq:rec_power}
  A_{n}^{k}=-\frac{\sqrt{(n+1)D}}{n}\partial_{x}A_{n+1}^{k-1}-\sqrt{\frac{D}{n}}\partial_{x}A_{n-1}^{k-1} + \frac{\partial_{x}\left(\partial_{x}\phi A_{n}^{k-2}\right)}{n}+\frac{T}{n}\partial_{xx}A_n^{k-2}\ .
\end{eqnarray}
Taking $k=n$ in \eref{eq:rec_power} and using that $A_{n}^{j}=0$ for $j\leq n$ yields the expression of $A_n^{n}$ as a function of $A_0^0$:
\begin{equation}
  \label{eq:rec_diagonal}
  A_n^{n}=-\sqrt{\frac{D}{n}}\partial_x A^{n-1}_{n-1} = (-1)^n \frac{D^{n/2}}{\sqrt{n!}}\partial_x^n A_0^0\ .
\end{equation}
Using expression \eref{eq:A_1_full} for $k=1$ and expression \eref{eq:rec_diagonal} for $n=1$, we obtain a closed equation on $A_0^0$:
\begin{equation}
  \label{eq:constrain_eq_measure}
  \partial_x\phi\ A_0^0 + (T+D)\partial_x A_0^0 =-b_1\sqrt{D}\ .
\end{equation}
Since $A_0^0$ corresponds to the equilibrium stationary measure when $\tau=0$ we must have
\begin{equation}
  \label{eq:eq_measure}
  \int_{-\infty}^{+\infty} \cP_s(x,v)|_{\tau=0}\ dv =A_0^0=c_{0}\ e^{-\frac{\phi}{T+D}}\ ,
\end{equation}
with $c_0$ fixed by normalization:
\begin{equation}
  \label{eq:c0}
  c_0=\left(\int_{-\infty}^{+\infty}e^{-\frac{\phi}{T+D}}dx\right)^{-1}\ .
\end{equation}
The constant $b_1$ is self-consistently fixed to zero such that \eref{eq:eq_measure} is a solution of \eref{eq:constrain_eq_measure}. We now set out to compute the next order correction $A_0^2$.
Applying \eref{eq:rec_power} for $n=1$ and $k=3$ gives:
\begin{equation}
  \label{eq:first_correction}
  A_{1}^{3}=-\sqrt{2D}\partial_{x}A_{2}^{2}-\sqrt{D}\partial_{x}A_{0}^{2}+\partial_{x}\left(\partial_{x}\phi A_{1}^{1}\right)+T\partial_{xx}A_1^{1}\ .
\end{equation}
In \eref{eq:first_correction}, we can use \eref{eq:rec_diagonal} to express $A_1^1$ and $A_2^2$ as a function of $A_0^0$ and \eref{eq:A_1_full} to express $A_1^3$ as a function of $A_0^2$.
We thus obtain a differential equation for $A_0^2$:
\begin{eqnarray}
  \label{eq:first_correction_closed}
  \fl
  \frac{\partial_x \phi }{T+D}A_0^2 + \partial_x A_0^2 &=&-\frac{D^2\partial_{x}^3 A_{0}^{0}}{T+D}+\frac{D\partial_{x}\left(\partial_{x}\phi \partial_x A_{0}^{0}\right)}{T+D}+\frac{TD\partial_{x}^3 A_0^{0}}{T+D} - \frac{b_3\sqrt{D}}{T+D}\ .
\end{eqnarray}
Using \eref{eq:eq_measure}, we can integrate \eref{eq:first_correction_closed} and determine the expression of $A_0^2$
\begin{eqnarray}
  \fl
  \label{eq:A_2}
  A_0^2 &=&\ c_0\ e^{-\frac{\phi}{T+D}}\left(\frac{D\partial_{xx}\phi}{T+D}-\frac{D\left(\partial_x\phi\right)^2}{2(T+D)^2}\right) + c_2\ e^{-\frac{\phi}{T+D}}- \frac{b_3\sqrt{D}}{T+D}e^{-\frac{\phi}{T+D}}\int_0^x e^{\frac{\phi}{T+D}}dx \ ,
\end{eqnarray}
where $c_0$ is defined in \eref{eq:c0}. Equation \eref{eq:A_2} involves two integration constants: $c_2$ and $b_3$.
While $c_2$ is found by normalization, requiring $\int_{-\infty}^{+\infty}A_0^2(x)dx=0$, $b_3$ is fixed by boundary conditions on $A_0^{2}$ as we shall see in the next sections.
The recursion can be iterated up to an arbitrary order in $\tau$ to find both the $A_0^{2k}$'s and the $A_i^{k}$'s for $i>0$. In addition to the previous constants $c_{2i}$ and $b_{2i+1}$ for $i<k$, which were determined for lower orders, $A_0^{2k}$ generically depends on two new integration constants : $c_{2k}$ and $b_{2k+1}$.
The former, $c_{2k}$, is found by requiring the normalization of $A_0^{2k}$ while the latter $b_{2k+1}$ is fixed by boundary conditions for $A_0^{2k}$.
For example, the differential equation on $A_0^4$ is found by applying \eref{eq:rec_power} for $(n=2,k=4)$ and $(n=1,k=5)$.
Its solution not only depends on $c_2$ and $b_3$, which were previously determined upon computing $A_0^2$, but also on two new integration constants : $c_{4}$ and $b_{5}$.
The constant $c_{4}$ is found by requiring normalization $\int^{+\infty}_{-\infty}A_0^{4}=0$ and $b_5$ is fixed by enforcing the correct boundary conditions for $A_0^4$.
While the explicit expressions of the $A_0^{2k}$ rapidly become cumbersome, their systematic derivation can be implemented with a software such as Mathematica \cite{Mathematica}.
For illustration purposes, we report the complete expression of $\cP_s(x,v)$, with its integration constants, up to the order $\tau^2$ in \eref{eq:full_distribution}.
We remark that \eref{eq:A_2} shares a common feature with the distribution of other active models \cite{van1984activation, woillez2020nonlocal}: it is non-local.
Indeed, a perturbation of the potential $\delta\phi(x)$ localized around position $x$ will affect the steady-state at position $x^{\prime}$ located far away from $x$.
This strongly differs from Boltzmann distribution and leads to intriguing phenomena, for example in bacterial suspensions \cite{galajda2007wall}.

\section{Confining potential: explicit computation and numerics}
\begin{figure}
  \centering
  \def\Y{2.0}
  \def\X{2.75}
  \begin{tikzpicture}

    \path (0,0)  node {\includegraphics[width=.45\columnwidth]{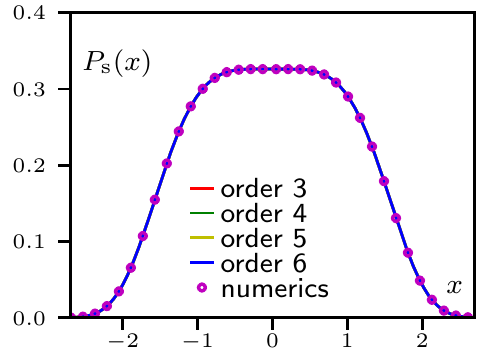}}
     (\X,\Y) node {{\bf{\large (a)}}};
    \path (7,0)  node {\includegraphics[width=.45\columnwidth]{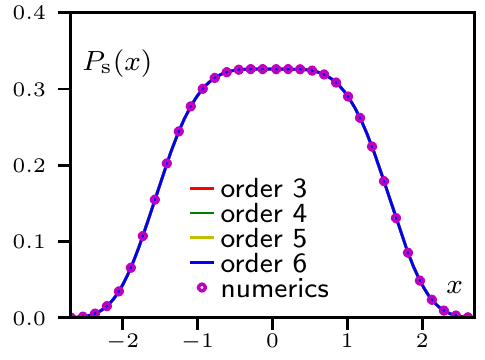}}
    (\X+7,\Y) node {{\bf{\large (b)}}};
    \path (0,-5.25)  node {\includegraphics[width=.45\columnwidth]{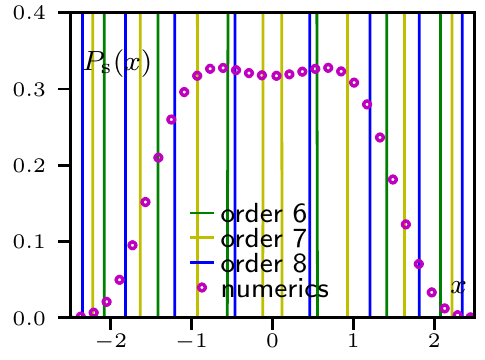}}
    (\X,\Y-5.25) node {{\bf{\large (c)}}};
    \path (7,-5.25)  node {\includegraphics[width=.45\columnwidth]{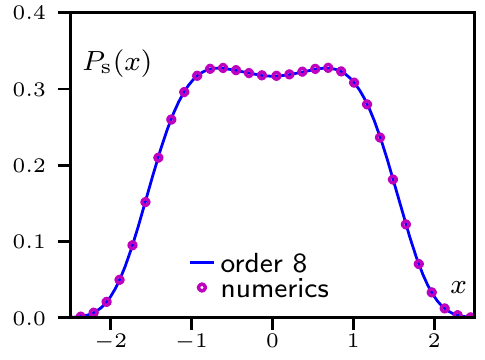}}
    (\X+7,\Y-5.25) node {{\bf{\large (d)}}};;

  \end{tikzpicture}
  \caption{Steady-state distribution of \eref{eq:langevin_1}-\eref{eq:langevin_2} in a confining potential $\phi(x) = x^4/4$
  {\bf Top:} For $\tau=0.01$, the finite truncation of \eqref{eq:marginal_confined} converges and agrees with the numerics {\bf (a)}. Its corresponding Borel resummation $B^r_8$ also coincides with simulation data  {\bf (b)}.
  {\bf Bottom:} For $\tau = 0.2$, the finite truncation of \eqref{eq:marginal_confined} is rapidly diverging {\bf (c)}. However, the Borel resummation $B^r_8$ accurately follows the data {\bf (d)}. Parameters : $D=T=1$, $dt=10^{-4}$, time $=10^8$.
  }
  \label{fig:measure_divergence}
  \vspace{-0.10cm}
\end{figure}

\begin{figure}
  \centering
  \includegraphics[width=.55\columnwidth]{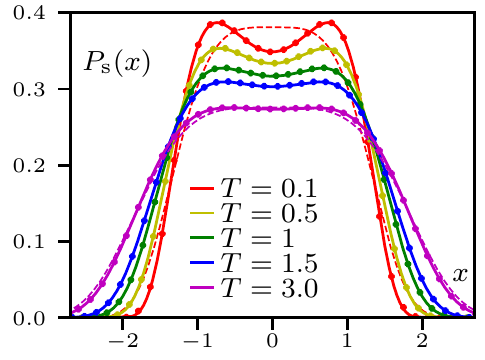}
  \caption{Steady-state distribution of \eqref{eq:langevin_1}-\eqref{eq:langevin_2} in a confining potential $\phi(x) = x^4/4$ for different values of $T$. Plain curves correspond to Borel resummations $B^r_8$ while symbols are obtained from numerical simulations of \eref{eq:langevin_1}-\eref{eq:langevin_2}. In dashed lines, we plot the Gibbs-Boltzmann distributions for the two limiting cases $T=0.1$ and $T=3.0$ to highlight the activity-induced deviation. The Borel resummation $B^r_8$ always fits the data accurately without any free parameters.
  Parameters: $\tau=0.2$, $D=1$, $dt=10^{-4}$, ${\rm time}=10^8$.
  }
  \label{fig:measure_pade}
\end{figure}

The marginal in space of $\cP_s(x,v)$ can be used to quantify how the steady-state distribution departs from the Boltzmann weight as $\tau$ increases :
\begin{equation}
  \label{eq:marginal_confined}
  \cP_s(x)=\int_{-\infty}^{+\infty}\cP_s(x,v)dv=A_0=\sum_k A_0^{2k}\tau^{k} \ .
\end{equation}
Here we consider the special case of a confining potential $\phi$, and we require that, for all $k\geq 1$,
\begin{eqnarray}
  \label{eq:integ_constrain_confined}
  \lim_{x\to\pm\infty}A_0^{2k}(x)&=&0 \\
  \label{eq:normalization_constrain_confined}
  \int_{-\infty}^{+\infty} A_0^{2k}(x)&=&0\ .
\end{eqnarray}
For a simple harmonic confinement, we note that the complete steady-state distribution $\cP_s(x,v)$ remains Gaussian and we report its expression in \ref{ap:harmonic}.
In the remainder of this paper, we will focus on the more general case of anharmonic potentials.
We remark that equation \eref{eq:integ_constrain_confined} imposes $b_{2k+1}=0$ for all $k\geq 1$ while \eref{eq:normalization_constrain_confined} fixes $c_{2k}$ for all $k\geq 1$.
The function $A_0^{2k}$ is then uniquely determined. For example, using \eref{eq:A_2} and the definition of $c_0$ \eref{eq:c0}, $A_0^2$ reads
\begin{eqnarray}
  \fl
  \label{eq:A_2_confined}
  A_0^2 =\ c_0\ e^{-\frac{\phi}{T+D}}\left(\frac{D\partial_{xx}\phi}{T+D}-\frac{D\left(\partial_x\phi\right)^2}{2(T+D)^2}\right) - \frac{3\ c_0^2\ D}{2(T+D)}\ e^{-\frac{\phi}{T+D}}\int_{-\infty}^{+\infty}\partial_{xx}\phi\ e^{-\frac{\phi}{T+D}}dx\ .
\end{eqnarray}
In expression \eref{eq:A_2_confined}, we can readily extract the first correction to the Gibbs-Boltzmann measure
\begin{equation}
  \fl
  \label{eq:correction_Gibbs}
  \frac{\cP_s(x)-c_0 e^{-\frac{\phi}{T+D}}}{c_0e^{-\frac{\phi}{T+D}}}=\tau\left[\frac{D}{T+D}\partial_{xx}\phi-\frac{3D}{T+D}\frac{\int_{-\infty}^{+\infty}\partial_{xx}\phi\  e^{-\frac{\phi}{T+D}}dx}{\int_{-\infty}^{+\infty} e^{-\frac{\phi}{T+D}}dx}\right]+o(\tau)\ .
\end{equation}
which reduces at $T = 0$ to the steady-state of an Active Ornstein Uhlenbeck (AOUP) particle \cite{FodorAOUP} at this order in $\tau$. The cumbersome expression of the full marginal in space $\cP_s(x)$ up to order $\tau^2$ is reported in (\ref{eq:full_marginal_confined}).
Note that our ansatz \eref{eq:scaling_A_0} rests on the hypothesis that $\cP_s (x)$ is an analytic function in $\tau^{1/2}$, which need not necessarily hold for an arbitrary potential. To check this hypothesis, we have to verify whether the series admits a finite radius of convergence.
We do this for a potential $\phi(x)=x^4/4$, at fixed $D$ and $T$ and for two different values of $\tau$. For $\tau=0.01$, we show in \Fref{fig:measure_divergence} that the truncation of \eref{eq:marginal_confined} to order $\tau^8$ is well-behaved and quantitatively agrees with the stationary distribution obtained numerically.
However, for $\tau=0.2$, \Fref{fig:measure_divergence} shows the successive orders of the truncation to be typical of asymptotic series: adding one order in $\tau$ increases the series by a larger amount than the sum of the previous terms, leading to wild oscillations.
While such a result seems disappointing, it does not mean that the full series fails in capturing the steady state.
Mathematically speaking, it only entails that the finite truncation yields a poor approximation of the full series and that more work should be carried out to extract physical behaviors.
To regularize our diverging truncated sequence, we resort to a Pad\'{e}-Borel summation method. We first introduce the Borel transform $B_N$ associated to \eref{eq:marginal_confined}:
\begin{equation}
  \label{eq:borel_transform}
  B_N(\tau)=\sum_{k=0}^{N}\frac{A_0^{2k}}{k!}\tau^{k}\ .
\end{equation}
The finite-$N$ truncation of the series \eref{eq:marginal_confined} is exactly recovered from its $N^{{\rm th}}$-Borel transform $B_N$ by applying a Laplace inversion :
\begin{equation}
  \label{eq:borel_inverse}
  \sum_{k=0}^{N}A^{2k}_0\tau^k = \int_0^{\infty}B_N(\omega\tau)e^{-\omega}d\omega\ .
\end{equation}
The Laplace inversion of expression \eref{eq:borel_transform} for $B_N$ indeed leads back to the divergent finite truncation that we wanted to regularize.
To avoid such a fate, one has to find a nonpolynomial approximation of $B_N(\tau)$ whose Taylor expansion coincides with the known terms in \eref{eq:borel_transform}.
In the Pad\'{e}-Borel method, it is achieved by approximating $B_N$ with a rational fraction $F_N=Q_N/R_N$, where $Q_N$ and $R_N$ are polynomials in $\tau$ of order $N/2$ chosen such that $B_N(\tau)=Q_N(\tau)/R_N(\tau)+o(\tau^N)$.
The Borel resummation at order $N$ of \eref{eq:marginal_confined}, $B^{r}_N$, is defined by replacing $B_N$ in \eref{eq:borel_inverse} by its Pad\'{e} approximant $F_N$ :
\begin{equation}
  B_N^{r} = \int_0^{\infty}\frac{Q_N(\omega\tau)}{R_N(\omega\tau)}e^{-\omega}d\omega\ .
\end{equation}
Finally, the series \eref{eq:marginal_confined} is formally obtained from the limit of $B^{r}_N$ when $N\rightarrow\infty$.
In this article, we estimate \eref{eq:marginal_confined} while keeping $N$ finite and we will not evaluate $B^r_{N}$ beyond $N=8$.
Interestingly, for $\tau=0.2$, while the truncated sequence of \eref{eq:marginal_confined} is divergent, its Borel resummation $B^r_8$ agrees quantitatively with numerical estimates of the steady-state distribution as shown in the bottom right corner of \Fref{fig:measure_divergence}.
In \Fref{fig:measure_pade}, we plot the Borel resummations $B^r_8$ and the corresponding numerics for different values of $T$.
When $T\ll D$, the dynamics \eref{eq:langevin_1}-\eref{eq:langevin_2} is strongly out-of-equilibrium and the probability density differs significantly from the Boltzmann weight with the presence of two humps.
When $T\gg D$, self-propulsion is washed out by thermal noise, the dynamics draws closer to equilibrium and the two humps of the distribution are smoothened out.
Note that the Borel resummation $B^r_8$ accurately fits the numerics without any free parameter.

\vspace{-0.10cm}

\section{Ratchet current: analytical formula and numerics}
\label{sec:current}
\begin{figure}
  \begin{tikzpicture}
    \path (-0.8,0) node {\includegraphics[width=.495\columnwidth]{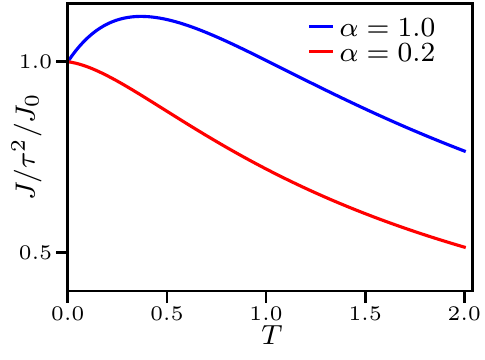}};
    \path (7.0,0) node {\includegraphics[width=.495\columnwidth]{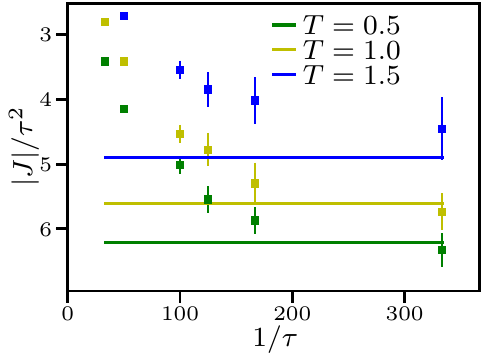}};
  \end{tikzpicture}
  \caption{Current $J$ induced by a ratchet potential $\phi(x) = \sin(\pi x/2) + \alpha \sin(\pi x)$ for different values of $T$ and $\alpha$.
  Plain curves correspond to prediction \eqref{eq:current_quant} while dots are numerical simulations with error bars given by the standard deviation. {\bf Left:} $J/\tau^2$ normalized by $J_0=J(T=0)$ as a function of $T$ for different values of $\alpha$.
  {\bf Right:} $J/\tau^2$ as a function of $1/\tau$ for $\alpha=1$.
  Parameters: $D=1$, $dt=15.10^{-4}$, ${\rm time}=5.10^8$.
  }
  \label{fig:current}
\end{figure}

An interesting signature of non-equilibrium dynamics is the ratchet mechanism by which asymmetric periodic potentials may lead to steady-state currents.
We consider here such a potential $\phi$ of period $L$ and we use our perturbative expansion to compute the steady-state current $J$, defined as
\begin{eqnarray}
  J&=&\langle \dot{x}\rangle \\
  &=&\int_{0}^{L}\int_{-\infty}^{\infty}\left(-\partial_x\phi+\frac{v}{\sqrt{\tau}}\right)\cP_s(x,v)dxdv \\
  \label{eq:J_non_simplified}
  &=&-\sum_{k\geq0}\int_0^{L}\partial_x\phi A_0^{2k}dx+\frac{\sqrt{ D}}{\sqrt{\tau}}\sum_{k\geq0}\int_0^{L}A_1^{2k+1}dx \\
  \label{eq:J_simplified}
  &=&\sum_{k\geq0} T\int^{L}_{0}\partial_x A_0^{2k}dx + L\sqrt{D}\sum_{k>0}b_{2k+1}\tau^{k}\ .
\end{eqnarray}
To go from \eqref{eq:J_non_simplified} to \eqref{eq:J_simplified}, we used the expression of $A_1^{2k+1}$ in \eqref{eq:A_1_full}.
We require the marginal in space $\cP_s(x)$ to be periodic, which entails $A_0^{2k}$ to be periodic for all $k\geq 0$ . The current $J$ then simplifies into :
\begin{equation}
  \label{eq:current_serie}
  J=L\sqrt{D}\sum_{k>0}b_{2k+1}\tau^{k}\ .
\end{equation}
While the $\{b_k\}$ all vanished in the previous section as a result of confinement \eqref{eq:integ_constrain_confined}, they do not for a periodic potential.
Indeed, the value of $b_k$ is fixed upon requiring the periodicity of $A_0^{k-1}$. Thus, different boundary conditions lead to different distributions, highlighting once again the nonlocal nature of the steady state.
We report the expression of the marginal in space $\cP_s(x)$ for a periodic potential up to order $\tau^2$ in \eref{eq:marginal_periodic}-\eref{eq:b5}.
Using it, we find that $Lb_5\tau^{2}$ is the first non-vanishing contribution to the current :
\begin{equation}
  \label{eq:current_quant}
  J=\frac{DL\tau^2}{2(T+D)}\frac{\int_0^{L} \phi^{(1)2}\phi^{(3)}dx}{\int_0^{L} e^{\frac{\phi}{T+D}}dx \int_0^{L} e^{-\frac{\phi}{T+D}}dx}+o(\tau^2)\ .
\end{equation}
The above formula reduces to the recently computed expression of $J$ for an AOUP particle when $T=0$ \cite{martin2020aoup}.
It is interesting to note that, as $T\rightarrow\infty$, $J$ always vanishes as $J\propto 1/T$.
Physically, when the thermal noise is much stronger than the self-propulsion, the nonequilibrium part of the dynamics becomes irrelevant and $J$ dies out.
However, as shown in the left part of \Fref{fig:current}, this intuitive picture is misleading at intermediate values of $T$. In this regime, the interplay between passive and active noises can, depending on the potential, make the current $J$ non-monotonic: ramping up the temperature might drive the particle further away from equilibrium.
In the right part of \Fref{fig:current}, we compare our quantitative prediction \eqref{eq:current_quant} with the results of numerical simulations for a potential $\phi(x) = \sin(\pi x/2) + \alpha\sin(\pi x)$ with $\alpha$ a constant.
We find quantitative agreement at small $\tau$ for $\tau < 0.01$, which confirms our conclusion in the previous section for the radius of convergence of our ansatz \eqref{eq:ansatz_1}.
Note that $J$ in \eqref{eq:current_serie} could also be regularized using Borel resummation to extend the quantitative range of agreement between theory and simulations to higher values of $\tau$, but we leave such a regularization for future works.

\vspace{-0.10cm}
\section{Entropy production rate}
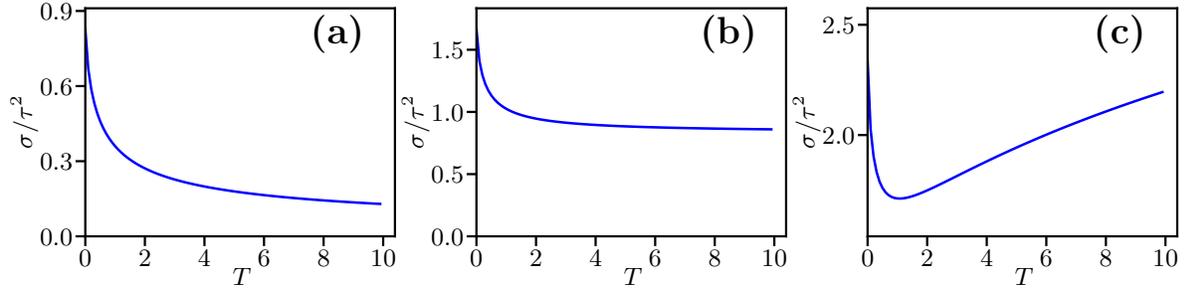
\begin{figure}
  \centering
  \def\Y{1.5}
  \def\X{1.8}
  \begin{tikzpicture}

    \path (0,0)  node {\includegraphics[width=.33\columnwidth]{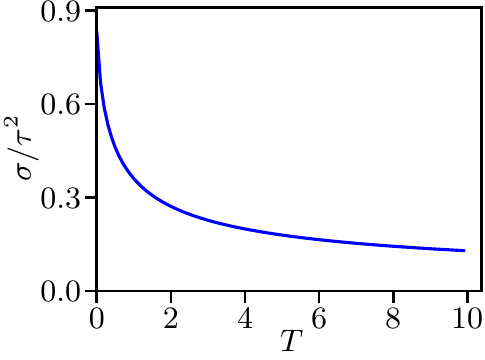}};
    \path (\X,\Y) node {{\bf{\large (a)}}};
    \path (5.2,0)  node {\includegraphics[width=.33\columnwidth]{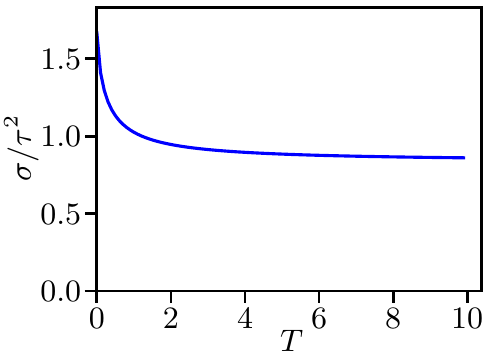}};
    \path (\X+5.2,\Y) node {{\bf{\large (b)}}};
    \path (10.4,0)  node {\includegraphics[width=.33\columnwidth]{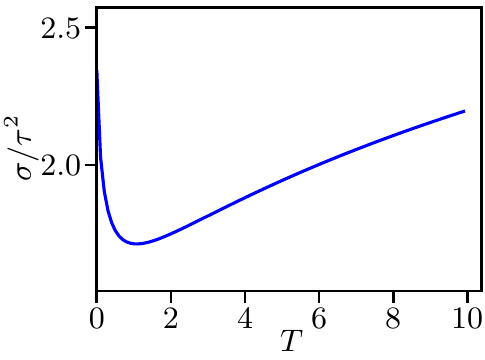}};
    \path (\X+10.4,\Y) node {{\bf{\large (c)}}};

  \end{tikzpicture}
  \vspace{-1cm}
  \caption{Entropy production rate of the process \eqref{eq:langevin_1}-\eqref{eq:langevin_2} divided by the persistence time squared in the small persistence time limit for different confining potential $\phi(x) = \lambda x^{2p}/2p!$ as a function of temperature at $D=1$ and $\lambda = 1$. For $p = 2$, the entropy production rate decreases as a function of $T$ and converges to $0$ at large $T$ {\bf (a)}. For $p=3$,  the entropy production rate decreases as a function of $T$ and converges to a non vanishing constant at large $T$ {\bf (b)}.
  For $p = 4$, the entropy production rate is a non monotonous function of $T$ and diverges at large $T$ {\bf (c)}.
  }
  \label{fig:entropy}
  \vspace{-0.3cm}
\end{figure}

Another signature of non-equilibrium processes is the existence of a non-zero entropy production rate $\sigma$, which is defined as the long time limit of the logarithm of the ratio between the probability of a trajectory and that of its time-reversed counterpart (to which we refer as "forward" and "backward" trajectories) divided by the duration of the trajectory. It thus measures the dynamics' irreversibility. Somehow counterintuitively, it has already been shown that $\sigma$ might exhibit a nonmonotonic behaviour when $\tau$ is varied \cite{dabelow2020irreversible,flenner2020active}.
In this part, in the same spirit,  we would like to assess the dependency of the entropy production rate on the temperature and explore its possible behaviours in different contexts.
More generally, the computation of $\sigma$ for the AOUP dynamics \eref{eq:langevin_1}-\eref{eq:langevin_2} remains a hot topic \cite{dabelow2019irreversibility,caprini2019entropy,FodorAOUP,flenner2020active} and it has triggered a debate about the parity of the self-propulsion $v$ under time-reversal \cite{caprini2018comment}.
Following \cite{caprini2019entropy}, we choose here to focus instead on the non Markovian process $x(t)$ obtained after integrating out the active degrees of freedom $v(t)$.
In this case, a trajectory over the time interval $[0,t_f]$ is solely defined as a set of positions $x(t)$ for $t \in [0,t_f]$ and its backward counterpart is given by the set of positions $\mathcal{R}x(t) = x(t_f-t)$.
After equilibration of the process $v(t)$, equations~\eqref{eq:langevin_1}-\eqref{eq:langevin_2} can be rewritten in position space only as
\begin{equation}
    \frac{d x}{d t} = - \phi^{\prime}(x) + \psi(t)\ ,
\label{eq:EOM_nonMarkov}
\end{equation}
where $\psi(t) = \sqrt{2T}\eta_1(t) + v(t)$ is a zero mean Gaussian noise with variance
\begin{equation}
    \left\langle \psi(t_1)\psi(t_2) \right\rangle = \frac{D}{\tau}\exp{\left(-\frac{|t_1-t_2|}{\tau}\right)} + 2T \delta(t_1 - t_2) = \Gamma(t_1-t_2)\ .
\label{eq:pathmeasure}
\end{equation}
To derive $\sigma$, we use a path-integral formalism. Since $\psi$ is Gaussian, we obtain the steady-state Itō probability of a trajectory over the time interval $[0,t_f]$ as:
\begin{equation}
    \mathcal{P}[{x(t)}] \propto P_s(x(0)) \exp{\left(-\frac{1}{2}\int_0^{t_f}\int_0^{t_f} d t_1 \, d t_2 \, \mathcal{S}[\dot{x},x]\right)}\ ,
    \label{eq:path_proba}
\end{equation}
with the action
\begin{equation}
\mathcal{S}[\dot{x},x] = \left[\dot{x}(t_1) + \phi^{\prime}(x(t_1))\right] \Gamma^{-1}(t_1-t_2)\left[\dot{x}(t_2) + \phi^{\prime}(x(t_2))\right]\ ,
\end{equation}
and where $P_s$ is the stationary state probability distribution and $\Gamma^{-1}(t)$ the functional inverse of the noise time correlation. It writes
\begin{equation}
  \Gamma^{-1}(t) = \frac{1}{2T}\delta(t) - \frac{G(t)}{\tau}\ ,
\end{equation}
with
\begin{equation}
    G(t) = \frac{D}{4 T^2}\sqrt{\frac{T}{D+T}}\exp{\left(-\sqrt{\frac{D+T}{T}}\frac{|t|}{\tau}\right)}\ .
\end{equation}
By definition, the entropy production rate $\sigma$ over a path $x(t)$ is given by
\begin{equation}
  \label{eq:entropy_def}
  \sigma = \lim_{t_f\to\infty}\frac{1}{t_f}\int_0^{t_f}\int_0^{t_f} d t_1 \, d t_2 \, \frac{1}{2}\left(\mathcal{S}\left[\dot{\mathcal{R}x},\mathcal{R}x\right]-\mathcal{S}\left[\dot{x},x\right]\right)\ .
\end{equation}
with $\mathcal{R}x$ the reverse path. Note that in \eref{eq:entropy_def}, even terms under time reversal cancels while exact derivatives yield no contribution to $\sigma$ in the limit $t_f\rightarrow \infty$.
Taking into account these simplifications, as well as the ergodicity of the dynamics that allows us to replace long-time averages by dynamical ensemble averages, we obtain the entropy production rate  as
\begin{equation}
  \label{eq:entropy_production_3}
  \sigma = - 2 \int_{-\infty}^{+\infty}\Gamma^{-1}(t)\left\langle\dot{x}(0)\phi^{\prime}(x(t))\right\rangle dt \, ,
\end{equation}
where the stochastic integral is now understood in the Stratonovich scheme and the average is computed using \eqref{eq:path_proba}.
Note that formula \eref{eq:entropy_production_3} for $\sigma$ is general and extends to any additive SDE with Gaussian colored noise. Eventually, as the local part of the kernel does not contribute to the entropy production rate, $\sigma$ expresses as
\begin{equation}
  \label{eq:entropy_production_4}
  \sigma = \frac{2}{\tau} \int_{-\infty}^{+\infty}G(t)\left\langle\dot{x}(0)\phi^{\prime}(x(t))\right\rangle dt \ .
\end{equation}
So far, the entropy production rate \eref{eq:entropy_production_3} involves two-time correlation functions, and our approach will be to reduce it to averages taken from the steady-state distribution computed in Section \ref{sec:pdf}.
To this aim, we use the particle displacement as a small-$\tau$ expansion parameter.
Indeed, over times of order $\tau$, for which the kernel $G(t)$ is non-vanishing, we have $x(t)-x(0)\sim\sqrt{\tau}$.
The details of this expansion are given in \ref{app:entropy}. In particular, \eref{eq:entropy_production_4} leads to the following expansion of $\sigma$
\begin{equation}
    \sigma = \frac{2}{\tau} \sum_{n=2}^{+\infty} \frac{1}{n!} \int_{0}^{+\infty} d t \, G(t) \left\langle \dot{x}(0) \phi^{(n+1)}(x(0)) \left[x(-t) - x(0)\right]^n \right\rangle\ .
\label{eq:vanishing_harmonic}
\end{equation}
where the discretization is of the Stratonovich type and where $\phi^{(k)}$ is the $k$-th derivative of $\phi$. In agreement with \cite{dadhichi2018origins}, \eqref{eq:vanishing_harmonic} allows us to show that additive SDEs with Gaussian colored noise have vanishing entropy production rates when the potential is harmonic.
Moreover, as shown in \ref{app:entropy}, the equation of motion \eqref{eq:EOM_nonMarkov} can be integrated recursively in powers of $\tau$ to yield a series expansion in $\tau^{1/2}$ of $\sigma$. Our main result is the first non-vanishing order in $\tau$ of this expansion
\begin{equation}
  \label{eq:entrop_prod_final}
  \sigma = D \tau^2 H\left(\frac{T}{D}\right) \frac{\int_{-\infty}^{+\infty}\phi^{(3) 2}\ e^{-\frac{\phi}{T+D}}dx}{\int_{-\infty}^{+\infty}e^{-\frac{\phi}{T+D}}dx}+O(\tau^{\frac{5}{2}})\ ,
\end{equation}
where the function $H$ is given by
\begin{equation}
  H(x)=\frac{4\sqrt{\frac{x}{x+1}}+x\left(4\sqrt{\frac{x}{x+1}}+2\right)+1}{8\sqrt{x(x+1)}+2x\left(6x+6\sqrt{x(x+1)}+7\right)+2}\ .
\end{equation}
When $T\rightarrow 0$, the entropy production rate \eref{eq:entrop_prod_final} brings us back to the expected findings of \cite{FodorAOUP} for an AOUP particle. Furthermore, in a system endowed with periodic boundary conditions at $-L$ and $+L$, the entropy production rate vanishes as $1/T$ at large temperature and
\begin{equation}
 \sigma \simeq \frac{D^2 \tau^2}{4T} \frac{\int_{-L}^{+L}\phi^{(3) 2}dx}{2L}\ .
\end{equation}
Physically, this supports the idea that thermal noise is washing out activity and nonequilibrium signatures.
However, this intuitive picture is challenged by the rich behavior of $\sigma$ with $T$, which strongly depends on the nature of $\phi$ and need not be a monotonic decreasing function.
For an unbounded system in a confining potential, the entropy production rate might even diverge at high temperature: increasing $T$ might thus drive the system further away from equilibrium.
In order to illustrate this idea, let us assume that $\phi(x) = \lambda x^{2p}/2p!$ with $p$ an integer great than 1. For $T \gg D$,
\begin{eqnarray}
\frac{\int_{-\infty}^{+\infty}\phi^{(3) 2}\ e^{-\frac{\phi}{T+D}}dx}{\int_{-\infty}^{+\infty}e^{-\frac{\phi}{T+D}}dx} & \sim \lambda \frac{(2p)!}{(2p-3)!}\frac{\int_{-\infty}^{+\infty}x^{4p-6}\ e^{-\lambda\frac{x^{2p}}{T}}dx}{\int_{-\infty}^{+\infty}e^{-\lambda\frac{x^{2p}}{T}}dx} \\ & \propto T^{2-3/p}\ ,
\end{eqnarray}
which shows that the entropy production rate behaves at high $T$ as
\begin{equation}
\sigma \propto T^{1-3/p}\ .
\end{equation}
As $T \to \infty$, it thus goes to $0$ for $p = 2$ and diverges for $p > 3$ as the particle explores steeper regions of the potential.
In Figure \ref{fig:entropy}, we plot $\sigma/\tau^2$ in the $\tau \to 0$ limit, as given by \eqref{eq:entrop_prod_final}, as a function of temperature in the three potentials characterized by $p=2$, $p=3$ and $p=4$ and for $D = 1$ and $\lambda = 1$.
Depending on the potential, it shows the rich phenomenology displayed by $\sigma$ when $T$ is varied: monotonic decrease or non-monotonicity, divergence or decay at high temperature...

\section{Conclusion}
We developed theoretical insights for an AOUP subjected to an additional Brownian noise \eqref{eq:langevin_1}-\eqref{eq:langevin_2}.
First, we devised a recurrence scheme allowing us to compute its stationary ditribution at an arbitrary order in $\tau^{1/2}$.
We then used this result to derive quantitative expressions for activity-induced phenomena such as the emergence of current and the entropy production rate.
We find that the interplay between passive and active noises produces a rich phenomenology for these non-equilibrium signatures when T is varied: monotonic, non-monotonic, diverging or decaying behaviors.
The intuitive picture of a passive noise hindering activity is thus challenged in many cases where switching on translational diffusion instead drives the particle further away from equilibrium.
As an alternative to our derivation, it is possible to obtain the marginal in space $\cP_s(x)$ up to order $\tau^2$ by using a Markovian approximation for the evolution operator: such a method has been developed in parallel to this article \cite{YongJooCom}.
We further remark that our perturbative approach can be generalized to more complex dynamics than the Brownian case.
Let us consider a stochastic dynamics $\mathcal{S}$ whose corresponding Fokker-Planck operator is $\cL_{\mathcal{S}}$. Adding an OU noise of amplitude $D$ to $\mathcal{S}$ amounts to adding $\cL_1 + \cL_2$ to $\cL_{\mathcal{S}}$.
The starting point for our perturbative expansion, $A_0^0$, is defined by the equation $(\cL_{\mathcal{S}} + D \partial_{xx})A_0^0 = 0$. As long as $A_0^0$ defined this way is known analytically, the recursion can be carried out and the effect of the OU noise can be taken into account perturbatively in $\tau$.
For example, using our method, one could try to assess the effect of a colored noise on an underdamped Langevin dynamics.
\ack
We thank Julien Tailleur, Fr\'{e}d\'{e}ric van Wijland and Cesare Nardini for insightful discussions, continuous support as well as enlightening feedback.
\appendix
\section{Full steady-state distribution}
\label{app:SS_distribution}
In this appendix, we report the steady-state probability density $\mathcal{P}_s(x,v)$ up to order $\tau^2$.
\begin{eqnarray}
  \fl
  \nonumber
  e^{\frac{\phi}{T+D}}\mathcal{P}_s(x,v) = c_0 + \sqrt{\tau} P_1(v)\frac{c_0 \sqrt{D} \phi^{(1)}(x)}{T+D}+\tau P_0(v) \bigg{[}-\frac{c_0 D \phi^{(1)2}}{2(T+D)^2}+c_2+\frac{Dc_0\phi^{(2)}}{T+D}\\
  \nonumber
  -\frac{\sqrt{D}b_3}{T+D}\int_0^x e^{\frac{\phi(z)}{T+D}}dz\bigg{]}+\tau^{\frac{3}{2}}\bigg{[}\frac{P_3(v)c_0 D^{\frac{3}{2}}}{\sqrt{6}}\bigg{(}\frac{\phi^{(1)3}}{(T+D)^3}-\frac{3\phi^{(1)}\phi^{(2)}}{(T+D)^2}+\frac{\phi^{(3)}}{T+D}\bigg{)}\\
  \nonumber
  +P_1(v)\bigg{(}\frac{b_3 D e^{\frac{\phi}{T+D}}}{T+D} -\frac{\phi^{(1)}b_3 D}{(T+D)^2}\int_0^x e^{\frac{\phi(z)}{T+D}}dz+\frac{\sqrt{D}c_2\phi^{(1)}}{T+D}-\frac{c_0 D^{\frac{3}{2}}\phi^{(1)3}}{2(T+D)^3}\\
  \nonumber
  +\frac{c_0\sqrt{D}(D^2-T^2)\phi^{(1)}\phi^{(2)}}{(T+D)^3}+\frac{c_0\sqrt{D}T\phi^{(3)}}{T+D}\bigg{)}\bigg{]}+\tau^2 P_0(v)\bigg{[}\frac{c_2 D\phi^{(2)}}{T+D}-\frac{c_2 D \phi^{(1)2}}{2(T+D)^2} \\
  \nonumber
  -\frac{D^{\frac{3}{2}}b_3}{(T+D)^2}\int_0^x e^{\frac{\phi(z)}{T+D}}\phi^{(2)}(z)dz +\frac{c_0 D^2}{8(T+D)^4}\phi^{(1)4}-\frac{c_0 D(D-T)\phi^{(1)2}\phi^{(2)}}{2(T+D)^3}\\
  \nonumber
  + \frac{b_3 D^{\frac{3}{2}}}{(T+D)^3}\int_0^x \left(\int_0^s e^{\frac{\phi(z)}{T+D}}dz\right)\left(\phi^{(1)}(s)\phi^{(2)}(s)-(T+D)\phi^{(3)}(s)\right)ds +c_4\\
  \nonumber
  +\frac{Dc_0}{2(T+D)}\int_0^x \phi^{(1)2}(z)\phi^{(3)}(z)dz -\frac{Dc_0(D+2T)}{(T+D)^2}\phi^{(3)}\phi^{(1)}-\frac{\sqrt{D}b_5}{T+D}\int_0^x e^{\frac{\phi(z)}{T+D}}dz\\
  \label{eq:full_distribution}
  +\frac{Dc_0(D-2T)\phi^{(2)2}}{4(T+D)^2}+\frac{Dc_0(D+2T)\phi^{(4)}}{2(T+D)}\bigg{]}\ .
\end{eqnarray}
In \eref{eq:full_distribution}, $c_0$ is defined by \eref{eq:c0} while $c_2$, $c_4$, $b_3$ and $b_5$ are integration constants whose expressions must be adapted to the boundary conditions.
For a confining potential, $\cP_s(x,v)$ must vanish for $x\rightarrow\pm\infty$, and thus $b_3=b_5=0$ yielding the following spatial distribution :
\begin{eqnarray}
  \fl
  \nonumber
  e^{\frac{\phi}{T+D}}\mathcal{P}_s(x) = c_0 +\tau\bigg{[}-\frac{c_0 D \phi^{(1)2}}{2(T+D)^2}+c_2+\frac{Dc_0\phi^{(2)}}{T+D}\bigg{]}
  +\tau^2 \bigg{[}\frac{c_2 D\phi^{(2)}}{T+D}-\frac{c_2 D \phi^{(1)2}}{2(T+D)^2} +c_4\\
  \nonumber
  +\frac{c_0 D^2}{8(T+D)^4}\phi^{(1)4}-\frac{c_0 D(D-T)\phi^{(1)2}\phi^{(2)}}{2(T+D)^3}+\frac{Dc_0}{2(T+D)}\int_0^x \phi^{(1)2}(z)\phi^{(3)}(z)dz\\
  \label{eq:full_marginal_confined}
  -\frac{Dc_0(D+2T)}{(T+D)^2}\phi^{(3)}\phi^{(1)}
  +\frac{Dc_0(D-2T)\phi^{(2)2}}{4(T+D)^2}+\frac{Dc_0(D+2T)\phi^{(4)}}{2(T+D)}\bigg{]} \ .
\end{eqnarray}
The integration constants $c_2$ and $c_4$ are finally found by normalization, requiring $\int_{-\infty}^{+\infty}  \cP_s(x)dx=1$ at every order in $\tau$.\\
For a periodic potential of period $L$, the spatial distribution must respect $\cP_s(x+L)=\cP_s(x)$. This condition implies $b_3=0$, but $b_5\neq 0$ and $\cP_s$ reads :
\begin{eqnarray}
  \fl
  \nonumber
  e^{\frac{\phi}{T+D}}\mathcal{P}_s(x) = c_0 +\tau\bigg{[}-\frac{c_0 D \phi^{(1)2}}{2(T+D)^2}+c_2+\frac{Dc_0\phi^{(2)}}{T+D}\bigg{]}
  +\tau^2 \bigg{[}\frac{c_2 D\phi^{(2)}}{T+D}-\frac{c_2 D \phi^{(1)2}}{2(T+D)^2} +c_4\\
  \nonumber
  +\frac{c_0 D^2}{8(T+D)^4}\phi^{(1)4}-\frac{c_0 D(D-T)\phi^{(1)2}\phi^{(2)}}{2(T+D)^3}+\frac{Dc_0}{2(T+D)}\int_0^x \phi^{(1)2}(z)\phi^{(3)}(z)dz\\
  \nonumber
  -\frac{Dc_0(D+2T)}{(T+D)^2}\phi^{(3)}\phi^{(1)}
  +\frac{Dc_0(D-2T)\phi^{(2)2}}{4(T+D)^2}+\frac{Dc_0(D+2T)\phi^{(4)}}{2(T+D)}\\
  \label{eq:marginal_periodic}
  -\frac{\sqrt{D}b_5}{T+D}\int_0^x e^{\frac{\phi(z)}{T+D}}dz\bigg{]}\ ,
\end{eqnarray}
with $b_5$ given by
\begin{eqnarray}
  \label{eq:b5}
  b_5 =\frac{D}{2(T+D)}\frac{\int_0^{L} \phi^{(1)2}\phi^{(3)}dx}{\int_0^{L} e^{\frac{\phi}{T+D}}dx \int_0^{L} e^{-\frac{\phi}{T+D}}dx}\ .
\end{eqnarray}
Once again, $c_2$ and $c_4$ are then found by normalization.
Note that in expression \eref{eq:full_distribution}, $v$ corresponds to the rescaled variable $\tilde{v}$. To get the exact steady-state distribution associated to \eref{eq:langevin_1}-\eref{eq:langevin_2}, one thus has to make the change of variable $v\rightarrow \sqrt{\tau} v$.
\section{Harmonic potential}
\label{ap:harmonic}
We report hereafter the steady-state distribution for the special case of a harmonic potential $\phi(x)= \kappa x^2/2$
\begin{eqnarray}
  \label{eq:pstat_harmo}
  \cP_s(x,v) = \frac{\sqrt{4ab-c^2}}{2\pi}e^{-a x^2-b v^2 + cvx}\ ,
\end{eqnarray}
with the constants $a$, $b$ and $c$ defined as :
\begin{eqnarray}
  \label{eq:harmonic_potential_app}
  \fl
  a &=&\frac{\kappa(1+\kappa\tau)^2}{2(D+T(1+\kappa\tau)^2)}\quad b=\frac{D(1+\kappa\tau)+T(1+\kappa\tau)^2}{2D(D+T(1+\kappa\tau)^2)}\quad c=\frac{\kappa\sqrt{\tau}(1+\kappa\tau)}{D+T(1+\kappa\tau)^2}\ .
\end{eqnarray}
Note that in expression \eref{eq:pstat_harmo}, $v$ corresponds to the rescaled variable $\tilde{v}$.
To get the exact steady-state distribution associated to \eref{eq:langevin_1}-\eref{eq:langevin_2}, one thus has to replace $v$ with $\sqrt{\tau} v$ in \eref{eq:pstat_harmo} and to multiply \eref{eq:pstat_harmo} by $\sqrt{\tau}$.
At $T=0$, the distribution \eref{eq:pstat_harmo} corresponds to the result obtained in \cite{szamel2014self}.

\section{Computing the entropy production rate}
As shown in \eqref{eq:entropy_production_4}, the entropy production rate can be expressed as
\begin{equation}
    \sigma = \frac{2}{\tau}\left\langle \int_{-\infty}^{+\infty}d t \,  G(t) \,  \dot{x}(0) \, \phi^{\prime}(x(t)) \right\rangle\ .
\label{eq:entropy_production}
\end{equation}
The small $\tau$ expansion of \eqref{eq:entropy_production} is obtained by expanding it in powers of the particle displacement. In order to make this expansion in $\tau$ more explicit, we rescale time as $s = t/\tau$ and active force as $\hat{v} = v\sqrt{\tau}$. The entropy production rate then writes
\begin{equation}
    \sigma = \frac{2}{\tau}\left\langle \int_{-\infty}^{+\infty}d s \,  \hat{G}(s) \,  \frac{d x}{d s}(0) \, \phi^{\prime}(x(s)) \right\rangle\ ,
\end{equation}
with
\begin{equation}
    \hat{G}(s) = \frac{D}{4 T^2}\sqrt{\frac{T}{D+T}}\exp{\left(-\sqrt{\frac{D+T}{T}}|s|\right)}\ ,
\end{equation}
and the path measure $\left\langle \dots \right\rangle $ corresponding now to the process
\begin{eqnarray}
    & \frac{d x}{d s} = - \tau \phi^{\prime}(x(s)) + \sqrt{\tau}\left(\hat{v}(s) + \sqrt{2T}\,\hat{\eta}_1(s)\right) \\
    & \frac{d \hat{v}}{d s} = - \hat{v} + \sqrt{2D}\,\hat{\eta}_2(s)\ ,
\label{eq:process}
\end{eqnarray}
where $\hat{\eta}_1(s)$ and $\hat{\eta}_1(s)$ are two independent Gaussian white noises. In order to keep notations simple we drop the hat in the following. We introduce the particle displacement during time $s$ as
\begin{equation}
    \Delta(s) = x(s) - x(0)\ .
\end{equation}
Hence, we have
\begin{equation}
    \sigma = \frac{2}{\tau}\int_{-\infty}^{\infty} d s\,G(s)\sum_{n=0}^{+\infty} \frac{1}{n!} \left\langle \dot{x}(0) \phi^{(n+1)}(x(0))\Delta(s)^n \right\rangle\ .
\label{eq:entropy_prod}
\end{equation}
As usual in stochastic calculus, the underlying discretization of the various expressions at hand is crucial. Therefore, throughout this appendix, and for the sake of clarity of the presentation, we will sometimes go back to the discrete limiting expressions. For instance, \eqref{eq:entropy_prod} is understood in the Stratonovich sense, \textit{i.e.} as the $\Delta t \to 0$ limit of the following discrete expression
\begin{equation}
  \fl
   \sigma = \frac{2}{\tau}\sum_{i = - \infty}^{+\infty} \Delta t \, G(i \Delta t) \sum_{n=0}^{+\infty} \frac{1}{n!} \left\langle \frac{\Delta x_0}{\Delta t}\,\phi^{(n+1)}\left(x_0 + \frac{\Delta x_0}{2}\right)\left(x_i - \left(x_0 + \frac{\Delta x_0}{2}\right)\right)^n \right\rangle\ ,
\end{equation}
with $\Delta x_i = x((i+1)\Delta t)-x(i \Delta t)$.
The first term of the series involves the Stratonovich average $\left\langle\dot{x}(0)\phi^{\prime}(x(0))\right\rangle$ and thus vanishes. We now focus on the second one that we denote by $\sigma_1$. We have
\begin{eqnarray}
    \nonumber \sigma_1 & = \frac{2}{\tau}\int_{-\infty}^{+\infty}d s \, G(s) \left\langle \dot{x}(0) \phi^{(2)}(x(0)) \Delta(s)\right\rangle \\
    \nonumber & = \frac{2}{\tau}\int_0^{+\infty}d s \, G(s) \left\langle \dot{x}(0) \phi^{(2)}(x(0))\left(\Delta(s)+\Delta(-s)\right)\right\rangle \\
    \nonumber & = \frac{2}{\tau}\int_0^{+\infty}d s \, G(s) \int_0^s d s^{\prime} \, \left\langle \dot{x}(0) \phi^{(2)}(x(0))\left(\dot{x}(s^{\prime}) - \dot{x}(-s^{\prime})\right)\right\rangle \\
    & = \frac{2}{\tau}\int_0^{+\infty}d s \, G(s) \int_0^s d s^{\prime} \, \left\langle \dot{x}(0)\dot{x}(s^{\prime}) \left(\phi^{(2)}(x(0)) - \phi^{(2)}(x(s^{\prime}))\right)\right\rangle\ ,
\label{eq:formula2}
\end{eqnarray}
where we have used time translation invariance in the steady state. The corresponding discretized expression writes
\begin{equation}
  \fl
  \sigma_1 = \frac{2}{\tau}\sum_{i=2}^{+\infty}\Delta t G(i\Delta t) \sum_{j=1}^{i-1} \left\langle\frac{\Delta x_0 \Delta x_j}{\Delta t} \left[\phi^{(2)}\left(x_0 + \frac{\Delta x_0}{\Delta t}\right) - \phi^{(2)}\left(x_i + \frac{\Delta x_i}{\Delta t}\right)\right]\right\rangle
\end{equation}
We now expand again \eqref{eq:formula2} in powers of the displacement, which gives
\begin{equation}
    \sigma_1 = - \frac{2}{\tau}\int_0^{+\infty}d s \, G(s) \int_0^s d s^{\prime} \, \sum_{n=1}^{+\infty} \frac{1}{n!} \left\langle \dot{x}(0) \phi^{(n+2)}(x(0)) \dot{x}(s^{\prime}) \Delta(s^{\prime})^n \right\rangle\ .
\end{equation}
In the Stratonovich discretization scheme we recognize a total derivative and we thus get
\begin{equation}
    \sigma_1 = - \frac{2}{\tau}\int_0^{+\infty}d s G(s) \, \sum_{n=1}^{+\infty}\frac{1}{(n+1)!}\left\langle \dot{x}(0) \phi^{(n+2)}(x(0)) \Delta(s)^{n+1}\right\rangle\ .
\end{equation}
Eventually, when plugged back in \eqref{eq:entropy_prod}, half of the terms cancel out and we obtain
\begin{equation}
    \sigma = \frac{2}{\tau} \sum_{n=2}^{+\infty} \frac{1}{n!} \int_{0}^{+\infty} d s \, G(s) \left\langle \dot{x}(0) \phi^{(n+1)}(x(0)) \Delta(-s)^n \right\rangle\ .
\label{eq:vanishing_harmonic_app}
\end{equation}
which is \eqref{eq:vanishing_harmonic} of the main text. Once more, \eqref{eq:vanishing_harmonic_app} should be understood in the Stratonovich sense, \textit{i.e.} as the continuous time limit of
\begin{equation}
  \fl
  \sigma = \frac{2}{\tau} \sum_{i=1}^{+\infty} \Delta t \, G(i \Delta t) \sum_{n=2}^{+\infty} \frac{1}{n!} \left\langle\phi^{(n+1)}\left(x_0 + \frac{\Delta x_0}{2}\right) \frac{\Delta x_0}{\Delta t}\left(x_{-i} - x_0 - \frac{\Delta x_0}{2}\right)^n\right\rangle\ .
\end{equation}
This first result justifies our claim that any additive Gaussian process in a harmonic potential has a vanishing entropy production rate.
Since $\left\langle \eta(0) x(-s) \right\rangle = 0$ for any $s>0 \,$, we are now in position to integrate out the thermal noise appearing in $\dot{x}(0)$.
This allows us to obtain an unambiguous continuous expression for the entropy production rate. Indeed, in \eqref{eq:vanishing_harmonic_app},
\begin{eqnarray}
     \nonumber &\left\langle \dot{x}(0) \phi^{(n+1)}(x(0)) \Delta(-s)^n \right\rangle \\ \nonumber
    & =  \left\langle \left(- \tau \phi^{\prime}(x(0)) + \sqrt{\tau} v(0) + \sqrt{2T\tau}\,\eta(0) \right)\phi^{(n+1)}(x(0))\left(x(-s) - x(0)\right)^{n} \right\rangle \\ \nonumber
    & = T \tau \left\langle \phi^{(n+2)}(x(0))\left(x(-s) - x(0)\right)^{n} - n \phi^{(n+1)}(x(0))\left(x(-s) - x(0)\right)^{n-1} \right\rangle \\ &+  \left\langle \left(- \tau \phi^{\prime}(x(0)) + \sqrt{\tau} v(0) \right)\phi^{(n+1)}(x(0))\left(x(-s) - x(0)\right)^{n} \right\rangle\ .
\end{eqnarray}
Note that the first term yields a telescopic sum. Then, using time translation invariance, one obtains the entropy production rate as
\begin{eqnarray}
  \fl
  \nonumber
  \sigma = \frac{2}{\tau}\left\langle\int_0^{+\infty} d s \, G(s) \left[\sum_{n=2}^{+\infty} \frac{(-1)^n}{n!}\left(-\tau \phi^{\prime}(x(s)) + \sqrt{\tau}v(s) \right)\phi^{(n+1)}(x(s))\Delta(s)^n \right] \right\rangle \\
    + \frac{2}{\tau}\left\langle\int_0^{+\infty} d s \, G(s) T \tau  \phi^{(3)}(x(s)) \Delta(s)\right\rangle\ .
\label{eq:entropy_prod_2terms}
\end{eqnarray}
\noindent So far this exact expression still involves two-time averages. In order to reduce the result to the evaluation of stationary state averages, we first expand again \eqref{eq:entropy_prod_2terms} in powers of $\Delta(s)$. The entropy production rate can thus be written as the sum of two contributions
\begin{equation}
    \sigma = \sigma_a + \sigma_b \ ,
\end{equation}
with the first one given by
\begin{eqnarray}
       \nonumber \sigma_a & =  2T \int_0^{+\infty} d s \, G(s) \left\langle \phi^{(3)}(x(s)) \Delta(s) \right\rangle \\   & = 2T \sum_{n=0}^{+\infty} \int_0^{+\infty} d s \, G(s)  \frac{\tau^{\frac{n+1}{2}}}{n!}\left\langle \phi^{(n+3)}(x(0))\left(\frac{\Delta(s)}{\sqrt{\tau}}\right)^{n+1}\right\rangle \ ,
\label{eq:sigmaa}
\end{eqnarray}
and the second one by
\begin{eqnarray}
  \fl
\sigma_b  &=  \int_0^{+\infty} d s \, G(s) \sum_{n=2}^{+\infty} \frac{2(-1)^n}{\tau n!} \left\langle \left(- \tau \phi^{\prime}(x(s)) + \sqrt{\tau} v(s) \right)\phi^{(n+1)}(x(s))\Delta(s)^{n}\right\rangle\ .
\label{eq:sigmab_0}
\end{eqnarray}
Taylor expanding \eref{eq:sigmab_0} around $x(0)$, we further express $\sigma_b$ as
\begin{eqnarray}
\nonumber
\fl
\sigma_b &= \sum_{n=2}^{+\infty}\sum_{p=0}^{+\infty} \frac{2(-1)^n}{p!\, n!} \tau^{\frac{n+p}{2}} \int_0^{+\infty} d s G(s) \left\langle \left.\partial_x^p\left[- \phi^{\prime}(x)\phi^{(n+1)}(x)\right]\right|_{x(0)}\left(\frac{\Delta(s)}{\sqrt{\tau}}\right)^{n+p}\right\rangle \\
\fl
&\quad + \sum_{n=2}^{+\infty}\sum_{p=0}^{+\infty} \frac{2(-1)^n}{p!\, n!} \tau^{\frac{n+p-1}{2}} \int_0^{+\infty} d s G(s) \left\langle \left.\left[v(s) \phi^{(n+1+p)}\right]\right|_{x(0)}\left(\frac{\Delta(s)}{\sqrt{\tau}}\right)^{n+p}\right\rangle\ .
\label{eq:sigmab}
\end{eqnarray}

\noindent Note that in \eqref{eq:sigmab}, the velocity is still evaluated at time s. This raises however no difficulty since the $v$ equation of motion can be integrated exactly as
\begin{equation}
    v(s) = v(0)e^{-s} + \sqrt{2D}e^{-s}\int_0^s d s^{\prime} \, e^{s^{\prime}}\eta_2(s^{\prime}) \ .
\end{equation}
Finally, in order to be able to use only stationary state averages when computing the entropy production rate, one needs to express $\Delta(s)$ as a function of $x(0)$. This is done by integrating the equation of motion recursively in powers of $\tau$,
\begin{equation}
    \frac{\Delta(s)}{\sqrt{\tau}} = -\sqrt{\tau} \int_0^s d s^{\prime} \, \phi^{\prime}(x(s^{\prime})) + \int_0^s d s^{\prime} \left(v(s^{\prime}) + \sqrt{2T}\eta_1(s^{\prime})\right)\ .
\label{eq:displacement_0}
\end{equation}
Applying \eref{eq:displacement_0} recursively in powers of $\tau$ allows us to compute $\Delta(s)$ up to order $\tau^{\frac{3}{2}}$
\begin{eqnarray}
  \fl
  \nonumber
  \frac{\Delta(s)}{\sqrt{\tau}} & =  - \sqrt{\tau} s \phi^{\prime}(x(0)) - \tau \int_0^s d s^{\prime} \frac{\phi^{\prime}(x(s^{\prime}))-\phi^{\prime}(x(0))}{\sqrt{\tau}} + \int_0^s d s^{\prime} \left(v(s^{\prime}) + \sqrt{2T}\eta_1(s^{\prime})\right) \\
  \fl
  \nonumber & =  \int_0^s d s^{\prime}\left( v(s^{\prime}) + \sqrt{2T}\eta_1(s^{\prime}) \right) - \sqrt{\tau} s \phi^{\prime}(x(0)) - \tau \phi^{(2)}(x(0)) \int_0^s ds^{\prime} \int_0^{s^{\prime}} d s^{\prime\prime} v(s^{\prime\prime}) \\
  \fl
  & \qquad - \tau \phi^{(2)}(x(0)) \int_0^s d s^{\prime} \int_0^{s^{\prime}} d s^{\prime\prime}\sqrt{2T}\eta_1(s^{\prime\prime}) + O(\tau^{3/2})
\label{eq:displacement}
\end{eqnarray}
where the above order in the expansion is enough to collect all terms of order $\tau^2$ in the entropy production rate.
Equation \eqref{eq:displacement} can then be plugged into \eqref{eq:sigmaa} and \eqref{eq:sigmab}. After averaging over the white noises $\eta_1(s)$ and $\eta_2(s)$, this allows us to obtain the entropy production rate, up to order $\tau^2$, solely expressed in terms of stationary state averages over both $x$ and $v$.
Using \eqref{eq:full_distribution}, we directly obtain \eqref{eq:entrop_prod_final} of the main text.
\label{app:entropy}
\section{Numerical methods}
To simulate dynamics \eref{eq:langevin_1}, we used a discretized Heun scheme while dynamics \eref{eq:langevin_2} was integrated exactly using Gillespie's method \cite{gillespie1996exact}. The obtained algorithm iterates as follows :
\begin{lstlisting}[mathescape=true]
$\mu$ = $\exp(-\rm{dt}/\tau)$;
$\sigma_x$ = $\sqrt{D(1-\mu^2)/\tau}$;
$Y_1$ = $\sqrt{2D\tau\left(\rm{dt}/\tau - 2(1-\mu) + 0.5(1-\mu^2)\right)-\tau D(1-\mu)^4/(1-\mu^2)}$;
$Y_2$ = $\sqrt{\tau D}(1-\mu)^2/\sqrt{1-\mu^2}$;
$T_1$ = $\sqrt{2T\rm{dt}}$;
Y = x = 0;
v	= $\sqrt{D/\tau}$*normal_distribution(0,1);

while(t < totaltime){
  $\eta_1$ = normal_distribution(0,1);
  $\eta_2$ = normal_distribution(0,1);
  $\eta_3$ = normal_distribution(0,1);
  Y = $\tau$*v*(1-$\mu$) + $Y_1$*$\eta_2$ + $Y_2$*$\eta_1$;
  v = v*$\mu$ + $\sigma_x$*$\eta_1$;
  $x_1$ = x - dt*$\partial_x\phi(x)$ + Y + $T_1$*$\eta_3$;
  x += Y + $T_1$*$\eta_3$ -0.5*dt*( $\partial_x\phi(x)$ + $\partial_x\phi(x_1)$ );
  t += dt;}
\end{lstlisting}
At step $(17)$, ${\rm x(t)}$ is stored in the variable ${\rm x}$.
The steady-state marginal in space of the distribution $\cP_s(x)$ was then obtained by recording the particle's position recurrently into an histogram.
The current $J$ was computed using the distance travelled by the particle divided by the duration of the simulation : the error bar on $J$ thus corresponds to the standard deviation.
Such a definition for the current was heuristically found to converge faster than computing $J=\langle -\partial_x\phi + v/\sqrt{\tau}\rangle$ with recurrent recordings.
\section*{References}
\bibliographystyle{iopart-num}
\bibliography{biblio_cam}
\end{document}